\documentclass[preprint, superscriptaddress, floatfix]{revtex4-2}
\usepackage{lineno}
\usepackage{graphicx}
\usepackage{float}
\usepackage{amssymb}
\usepackage{mathrsfs}
\usepackage{physics}
\usepackage{hyperref}
\usepackage{booktabs}
\usepackage[british]{babel}
\usepackage{caption}
\usepackage{bm}
\usepackage{xcolor}
\begin{document}

\title{Deterministic multi-phonon entanglement between two mechanical resonators on separate substrates}

\author{Ming-Han Chou} \thanks{These authors contributed equally to this work; Present address: AWS Center for Quantum Computing, Pasadena, CA 91125, USA}
\affiliation{Pritzker School of Molecular Engineering, University of Chicago, Chicago IL 60637, USA}
\affiliation{Department of Physics, University of Chicago, Chicago IL 60637, USA}

\author{Hong Qiao} \thanks{These authors contributed equally to this work.}
\affiliation{Pritzker School of Molecular Engineering, University of Chicago, Chicago IL 60637, USA}

\author{Haoxiong Yan}
\affiliation{Pritzker School of Molecular Engineering, University of Chicago, Chicago IL 60637, USA}

\author{Gustav Andersson}
\affiliation{Pritzker School of Molecular Engineering, University of Chicago, Chicago IL 60637, USA}

\author{Christopher R. Conner}
\affiliation{Pritzker School of Molecular Engineering, University of Chicago, Chicago IL 60637, USA}

\author{Joel Grebel} \thanks{Present address: Google, Santa Barbara, California 93117, USA}
\affiliation{Pritzker School of Molecular Engineering, University of Chicago, Chicago IL 60637, USA}

\author{Yash J. Joshi}
\affiliation{Pritzker School of Molecular Engineering, University of Chicago, Chicago IL 60637, USA}

\author{Jacob M. Miller}
\affiliation{Department of Physics, University of Chicago, Chicago IL 60637, USA}

\author{Rhys G. Povey}
\affiliation{Department of Physics, University of Chicago, Chicago IL 60637, USA}

\author{Xuntao Wu}
\affiliation{Pritzker School of Molecular Engineering, University of Chicago, Chicago IL 60637, USA}

\author{Andrew N. Cleland}
\email{anc@uchicago.edu}
\affiliation{Pritzker School of Molecular Engineering, University of Chicago, Chicago IL 60637, USA}
\affiliation{Center for Molecular Engineering and Material Science Division, Argonne National Laboratory, Lemont IL 60439, USA}

\maketitle

{\bf Mechanical systems have emerged as a compelling platform for applications in quantum information, leveraging recent advances in the control of phonons, the quanta of mechanical vibrations. Several experiments have demonstrated control and measurement of phonon states in mechanical resonators integrated with superconducting qubits \cite{OConnell2010, Chu2017, Satzinger2018, ArrangoizArriola2019}, and while entanglement of two mechanical resonators has been demonstrated in some approaches \cite{Riedinger2018, OckeloenKorppi2018,Kotler2021, Wollack2022,Luepke2024}, a full exploitation of the bosonic nature of phonons, such as multi-phonon entanglement, remains a challenge. Here, we describe a modular platform capable of rapid multi-phonon entanglement generation and subsequent tomographic analysis, using two surface acoustic wave resonators on separate substrates, each connected to a superconducting qubit. We generate a mechanical Bell state between the two mechanical resonators, achieving a fidelity of $\mathcal{F} = 0.872\pm 0.002$, and further demonstrate the creation of a multi-phonon entangled state (N=2 N00N state), shared between the two resonators, with fidelity $\mathcal{F} = 0.748\pm 0.008$. This approach promises the generation and manipulation of more complex phonon states, with potential future applications in bosonic quantum computing in mechanical systems. The compactness, modularity, and scalability of our platform further promises  advances in both fundamental science and advanced quantum protocols, including quantum random access memory \cite{Hann2019,Wang2024} and quantum error correction \cite{Chamberland2022}.}

Mechanical systems have significantly smaller footprints than existing circuit quantum electrodynamics (cQED) systems at similar frequencies \cite{ArrangoizArriola2019}, potentially long lifetimes \cite{MacCabe2020}, and a large number of accessible microwave-frequency modes \cite{Moores2018, Sletten2019, Chu2017, Chu2018}. Mechanical systems have been operated in the quantum limit \cite{OConnell2010, Chan2011}, with explorations of quantum information storage and processing \cite{Hann2019, Wallucks2020, Chamberland2022, MacCabe2020, Qiao2023} and quantum sensing \cite{Wollman2015, Mason2019,Huang2024}. Additional achievements include the quantum control of mechanical motion \cite{OConnell2010, Chu2017, Satzinger2018}, entanglement between macroscopic mechanical objects \cite{Palomaki2013, Riedinger2018, OckeloenKorppi2018,Kotler2021, Wollack2022}, coupling between surface acoustic waves (SAW) and qubits \cite{Gustafsson2014, Manenti2017, Noguchi2017, Moores2018, Bolgar2018}, the deterministic emission and detection of individual SAW phonons as well as phonon-phonon entanglement \cite{Bienfait2019, Bienfait2020, Dumur2021}, and the transmission of quantum information \cite{Bienfait2019, Bienfait2020, Dumur2021, Zivari2022, Zivari2022a}, among other demonstrations \cite{Delic2020, Shao2022, Zhang2018}. Mechanical systems have also been investigated as a platform for interconnecting microwave qubits with optical photons \cite{Bochmann2013, Andrews2014, Vainsencher2016, Peairs2020, Mirhosseini2020} and spin assemblies \cite{Whiteley2019}, with the potential for realizing long-distance quantum communication. Many of these advances have been enabled through the integration of superconducting qubits with mechanics, affording the quantum control of highly linear mechanical modes as well as straightforward quantum measurement. 

Here, we demonstrate the deterministic generation and distribution of multi-phonon entanglement between two physically separated mechanical modes. We use a modular architecture, in which the two mechanical resonators are fabricated on separate piezoelectric substrates and electrically coupled to a pair of superconducting qubits on a third, non-piezoelectric substrate. This design supports the generation of complex entangled states as well as straightforward quantum tomography, with potential applications in quantum sensing and high-precision measurements \cite{Degen2017, Carney2021, Goryachev2021, Schrinski2023,Linehan2024}. 
\begin{figure}[t]
\begin{center}
	\includegraphics[width=0.7\textwidth]{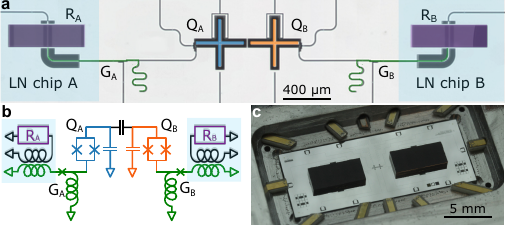}
	\caption{
    \label{fig1}
    {\bf Device layout, lumped circuit model, and optical micrographs.}
    \textbf{a,} False-colour optical micrograph of a device  identical to that used in the experiment. Two lithium niobate (LN) dies (light blue) each support one mechanical resonator $R_A$ and $R_B$ (left and right; purple), and are flip-chip bonded to a larger sapphire substrate, the latter including two qubits $Q_A$ (blue) and $Q_B$ (orange), their associated variable couplers $G_A$ and $G_B$ (green), and all readout resonators and control lines. \textbf{b,} Equivalent lumped-element circuit diagram. The air-gap inductive coupling between the sapphire wiring (green) and LN wiring (black) allows non-galvanic contacts between the LN dies and the sapphire substrate \cite{Satzinger2019}. \textbf{c,} Optical micrograph of assembled device. The two LN dies are $4.5\times 2$ $\text{mm}^2$, and the larger sapphire substrate is $15 \times 6~\text{mm}^2$. The $\sim 5~\mu$m air gap between the LN and sapphire dies is set by lithographically-defined epoxy standoffs \cite{Satzinger2019}. More details can be found in the Supplementary.
    }
\end{center}
\end{figure}

\begin{figure}[ht!]
\begin{center}
\includegraphics[width=0.7\textwidth]{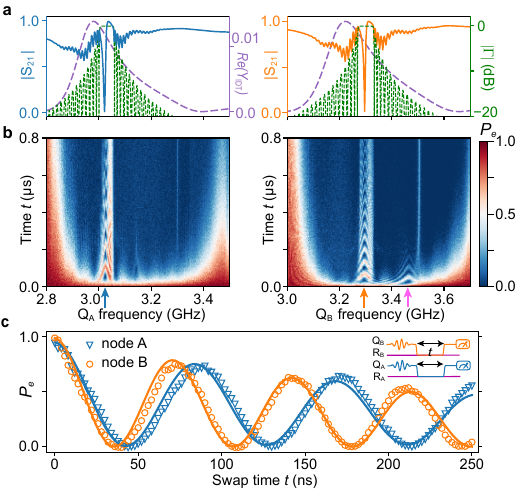}
\caption{
\label{fig2}
{\bf Device modeling and characterization.}
\textbf{a,} Numerically-calculated SAW resonator ($R_A$ and $R_B$) characteristics. Solid lines (blue and orange) show SAW resonator electrical transmission $|S_{21}|$ in linear scale (left) and green dashed lines indicate the acoustic mirror reflection coefficient $|\Gamma|$, with vertical axis in dB (right), together with the real part of the IDT admittance $Re(Y_{\text{IDT}})$ (S) (purple dashed line), all calculated using the coupling-of-modes (COM) model \cite{Morgan2010}. \textbf{b,} Qubit interaction with mechanical resonators, measured by monitoring either qubit's excited state probability $P_e$ (color scale) with time (vertical axis) while tuning the qubit frequency across the IDT bandwidth (horizontal axis). Interaction is measured separately for each qubit-resonator pair; response agrees well with the model using measured parameters. Blue and orange arrows indicate SAW resonant response, centered on the vacuum-Rabi exchanges of single quanta between the qubit and resonant mode. Pink arrow in right panel indicates $Q_A$ idle frequency. \textbf{c,} Simultaneous vacuum-Rabi swaps between each mechanical resonator and the associated qubit, with qubits set to the frequencies indicated by the blue and orange arrows in panel \textbf{b}. Solid lines are simulation results based on separate measurements of the mechanical lifetimes and qubit coherence times (See Extended Data Fig.~1). Inset shows pulse sequence (coupler control pulses not shown). 
}
\end{center}
\end{figure}

The experimental layout is shown in Fig~\ref{fig1}. Our device comprises two nodes, each node including a mechanical SAW resonator inductively coupled via a variable coupler \cite{Satzinger2018} to a frequency-tunable superconducting Xmon qubit \cite{Koch2007, Barends2013}. The two qubits are capacitively coupled to one another, supporting entangling gates. The two SAW resonators ($R_A$ and $R_B$) are fabricated on separate lithium niobate (LN) substrates, while the qubits ($Q_A$ and $Q_B$), couplers ($G_A$ and $G_B$), and their associated readout resonators and control lines, are fabricated on a sapphire substrate. The two LN dies are sequentially aligned and attached to the sapphire substrate, using a non-galvanic flip-chip assembly \cite{Satzinger2019}. Each mechanical resonator includes a central interdigitated transducer (IDT) and two acoustic mirrors, situated on either side of, and immediately adjacent to, the transducer. Each acoustic mirror is an array of two hundred 10 nm-thick parallel aluminium lines, forming a Bragg mirror with a $\sim$ 50 MHz-wide acoustic stop-band, centered on the respective mechanical resonator frequencies of 3.027 GHz ($R_A$) and 3.295 GHz ($R_B$). The free spectral range (FSR) of each acoustic resonator is designed to be slightly larger than mirror stop-band, confining a single acoustic mode in each resonator. By virtue of the piezoelectric response of the LN substrates, applying an electrical signal to either IDT generates symmetric, oppositely-directed surface acoustic waves, whose retro-reflection by the two mirrors forms a single Fabry-P\'erot resonance, generating a sympathetic electrical response at the corresponding IDT. In Fig.~\ref{fig2}\textbf{a} we show the calculated SAW resonator transmission, mirror stopband, and IDT admittance for each resonator, using their design parameters. Each superconducting qubit is coupled to its respective mechanical resonator via a variable coupler, whose coupling strength is controlled externally by magnetic flux bias of an rf-SQUID \cite{Satzinger2018}, with the coupler connected to the respective IDT through an air-gap inductive coupler. The three-die assembly is mounted in an aluminium box with wire-bond electrical connections and external magnetic shielding, operated in a dilution refrigerator with a base temperature of about 10 mK.

We first characterize each node by measuring the qubit-resonator interactions, measured as a function of time versus qubit frequency (see Fig.~\ref{fig2}\textbf{b}). Each qubit is initially excited to its $\ket{e}$ state by a tuned microwave $\pi$ pulse, following which the coupling between the qubit and resonator is turned on. When the qubit is tuned into resonance with the corresponding acoustic mode, qubit-resonator Rabi swaps give rise to the expected chevron patterns (Fig.~\ref{fig2}\textbf{b}). Outside the $\sim 50$ MHz acoustic mirror stop band, visible in Fig.~\ref{fig2}\textbf{a} (dashed green line), but within the IDT emission band of $\sim 600$ MHz (shown by the larger calculated admittance, dashed purple line), the qubit decays rapidly by acoustic emission that escapes through the mirrors. When the qubit is tuned outside the IDT emission band (left and right margins of either plot in panel \textbf{b}), the qubit lifetime increases rapidly due to the reduced phonon emission rate, due to the smaller admittance (purple dashed line in panel \textbf{a}). 

The system supports multiplexed Rabi swap measurements, shown in Fig.~\ref{fig2}\textbf{c}. Each qubit is set to its idle frequency of 3.245 GHz (3.557 GHz), a microwave $\pi$ pulse applied, and the qubits then tuned into resonance with their corresponding mechanical resonators while both variable couplers are turned on, yielding simultaneous parallel swaps. The swap times for nodes A and B are 42 ns and 35 ns, respectively. We next use the qubits to measure the mechanical resonator lifetimes at the single-phonon level, by swapping an excitation from the qubit into the resonator and then measuring the decay of the resulting one-phonon state as a function of time, for both nodes A and B. The resonators' energy relaxation times extracted from the measurements are $T_{1,A}^{m} = 380 \pm 8$ ns and $T_{1,B}^{m} = 270  \pm 3$ ns. The dephasing time for each resonator is then measured by exciting either qubit with a $\pi/2$ microwave pulse and swapping the qubit superposition state into the corresponding resonator, then measuring the decoherence time with a Ramsey fringe measurement \cite{Satzinger2018}. We find dephasing times of $T_{2,A}^{m} = 709 \pm 16$ ns and $T_{2,B}^{m} = 527 \pm 6$ ns, approximately twice the $T_{1}$ times (see Extended Data Fig.~1). The corresponding quality factors for the resonators are $Q_A \approx7200$ and $Q_B \approx 5600$, roughly twice the value for single-mode SAW resonator in Ref. \cite{Satzinger2018}. The improvement is possibly due to a modified SAW resonator geometry as well as more thorough surface cleaning of the LN substrates (see Supplementary). 

\begin{figure}[ht]
\begin{center}
	\includegraphics[width=0.7\textwidth]{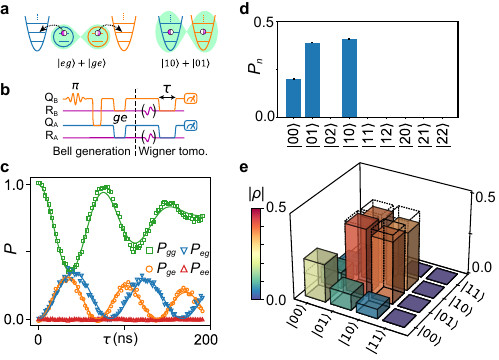}
	\caption{
    \label{fig3}
    {\bf Deterministic mechanical Bell state generation and tomography.} {\bf a,} Principle: A two-qubit Bell state $(|eg\rangle+|ge\rangle)/\sqrt{2}$ is generated in the qubits, then swapped coherently into the resonators, generating the entangled mechanical state $(|10\rangle+|01\rangle)/\sqrt{2}$. {\bf b,} Left of dashed line, pulse sequence to generate, then swap, a Bell state; right, Wigner tomography pulses, where the optional pulses for $R_A$ and $R_B$ indicate classical displacement pulses. {\bf c,} Joint qubit state probabilities with no displacement pulse; solid lines are fits, yielding joint resonator occupation probabilities in {\bf d}. {\bf e,} Density matrix from Wigner tomography of mechanical Bell state (solid colored bars), yielding a state fidelity $\mathcal{F} = 0.872\pm 0.002$ to the ideal Bell state (dashed bar outlines). The density matrix is reconstructed from tomography measurements. 
}
\end{center}
\end{figure}

We then use the qubits to prepare entangled mechanical resonator states, distributed across the two LN dies, as illustrated in Fig.~\ref{fig3}\textbf{a}. In Ref.~\cite{Wollack2022}, a two-resonator mechanical Bell state was prepared, then measured dispersively using a single-qubit Ramsey measurement. Here we use direct swaps between the resonators and their respective qubits for state analysis, using short pulse sequences with improved state fidelity. Following a protocol similar to Ref.~\cite{Wang2011} (pulse sequence shown in Fig.~\ref{fig3}\textbf{b}), we prepare a mechanical Bell state by exciting qubit $Q_A$ and performing a half-swap to qubit $Q_B$, generating a two-qubit Bell state $(|eg\rangle+|ge\rangle)/\sqrt{2}$. We then bring each qubit into resonance with its corresponding mechanical resonator, and turn on the variable couplers to perform full qubit-resonator swaps, ideally resulting in a dual-resonator Bell state $(|10\rangle+|01\rangle)/\sqrt{2}$. Note this method is not negatively impacted by the different swap times for nodes A and B (44.8 ns and 36.4 ns, respectively). To analyze the resulting entangled resonator state, coherent displacement pulses $\hat{D}_A$ and $\hat{D}_B$ are applied to each resonator, following which the qubits interact resonantly with their corresponding resonators, followed by simultaneous two-qubit state readout ~\cite{Hofheinz2009,Wang2011}. By varying the interaction time $\tau$, we can map out the two-qubit state probabilities $P_{gg}, P_{ge}, P_{eg}$ and $P_{ee}$ as a function of time, shown for zero displacement ($\hat{D}_A = \hat{D}_B = 0$) in panel \textbf{c}. These data show coherent swaps between the resonators and qubits while $P_{ee}$ remains zero, consistent with the expectation that only a single phonon is shared between the two resonators. Panel \textbf{d} shows the populations for the resonators, corresponding to the fit solid lines in \textbf{c}. By performing similar measurements with a total of $15 \times 15$ different combinations of displacement pulses $\hat{D}_A$ and $\hat{D}_B$ (see Supplementary), we reconstruct the Bell state using convex optimization. The resulting density matrix $|\rho|$ is shown in Fig.~\ref{fig3}\textbf{e}, with a fidelity $\mathcal{F} = \sqrt{\Tr(\rho_{\rm Bell}\cdot|\rho|)} = 0.872\pm 0.003$ to the ideal Bell state $\rho_{\rm Bell}$, close to our simulated result, which predicts a fidelity $\mathcal{F} = 0.92$ (see Supplementary). The infidelity is dominated by the resonator lifetime combined with a reduced qubit $T_1$ when each qubit is coupled to its resonator (see Supplementary and Extended Data Table.~1).

\begin{figure}[ht]
\begin{center}
	\includegraphics[width=0.7\textwidth]{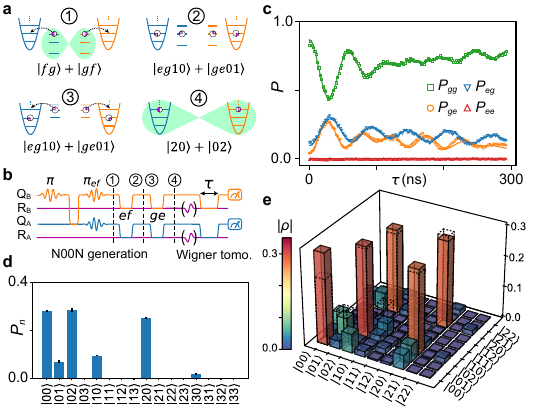}
	\caption{
    \label{fig4}
    {\bf Multi-phonon entanglement generation and tomography.} {\bf a,} Illustration of N00N ($N=2$) state generation process. We first generate an entangled qutrit state $(|fg\rangle+|gf\rangle)/\sqrt{2}$ ($|f\rangle$ is the qubit 2nd excited state), followed by a two-step swap from the qubits into the mechanical resonators, yielding a $(|20\rangle+|02\rangle)/\sqrt{2}$ N00N state. The corresponding pulse sequence is shown in {\bf b}; following state preparation, tomography is performed in a manner analogous to that for the Bell state tomography. {\bf c, d,} Qubit coincidence probability measurements and corresponding joint resonator population distribution. For the N00N $N=2$ state, the oscillations in $P_{eg}$ and $P_{ge}$ are approximately $\sqrt{2}$ faster than for the analogous Bell state measurements in Fig.~\ref{fig3}\textbf{b}; $P_{ee}$ remains zero as expected. {\bf e,} Density matrix resulting from Wigner tomography of multi-phonon entangled state, with a state fidelity $\mathcal{F} = 0.748\pm 0.008$ to the ideal $(|20\rangle+|02\rangle)/\sqrt{2}$ N00N state; measured density matrix $\rho$ is shown with solid color bars, while dashed outlines show the simulated result (see Supplementary).
    }
\end{center}
\end{figure}

We finally display the capabilities of this system for generating and measuring multi-phonon entangled states, doing so for an $N = 2$ N00N state shared between the two mechanical resonators. Our protocol is illustrated in Fig.~\ref{fig4}\textbf{a}, with the corresponding pulse sequence in panel \textbf{b}. We use a similar process to Fig.~\ref{fig3} to prepare a two-qubit Bell state $(|eg\rangle+|ge\rangle)/\sqrt{2}$, then use a microwave $\pi$ pulse to selectively excite each qubit to its second excited state $|f\rangle$, yielding the entangled qutrit state $(|fg\rangle+|gf\rangle)/\sqrt{2}$ (stage 1 in Fig.~\ref{fig4}\textbf{a}). We then tune each qubit's $f \leftrightarrow e$ transition into resonance with the corresponding resonator, and perform a full swap, resulting in the four-fold entangled state $(|eg10\rangle+|ge01\rangle)/\sqrt{2}$ (stage 2 in Fig.~\ref{fig4}\textbf{a}).  We note that during the $f \leftrightarrow e$ swap with the resonator, the $e \leftrightarrow g$ transition also falls inside the active SAW transducer bandwidth, but detuned from the resonator transition and outside the mirror bandwidth of $\sim 50$ MHz; the qubit $|e\rangle$ state thus decays in parallel by emitting unwanted, non-resonant phonons via the transducer, competing with the desired $f \rightarrow e$ transition. There is thus a trade-off between the qubit-resonator coupling strength and this unwanted phonon emission. This issue could be alleviated by reducing the IDT bandwidth so that $e \leftrightarrow g$ transition is outside IDT emission bandwidth due to qubit anharmonicity, e.g. by adding more IDT finger pairs. In our experiment, we carefully control both couplers to maximize the final N00N state fidelity. 

In the last step, each qubit's $e \leftrightarrow g$ transition is brought into resonance with its corresponding resonator, swapping the remaining qubit excitation into the resonator, ideally resulting in a final state $|gg\rangle \otimes (|20\rangle+|02\rangle)/\sqrt{2}$. This protocol can be extended to $(|N0\rangle+|0N\rangle)/\sqrt{2}$ states by iterating the first two steps. Alternatively, we can generate $(|N0\rangle+|0M\rangle)/\sqrt{2}$ N00M states, if one qubit is initially excited to its $|f\rangle$ state while the other qubit remains in $|e\rangle$. 

We analyze the final resonator state using Wigner tomography, similar to the Bell state analysis. The evolution of the two-qubit state probabilities for zero displacement are shown in Fig.~\ref{fig4}\textbf{c}; the corresponding joint resonator population distribution is shown in panel \textbf{d}. We see that the $P_{eg}$ and $P_{ge}$ oscillations are approximately $\sqrt{2}$ faster than the corresponding oscillations for the Bell state in Fig.~\ref{fig3}\textbf{b}, while the measured $P_{ee}$ remains zero, consistent with the expectation that detection of a phonon by one qubit precludes detection by the other qubit. Using a total of 261 different combinations of tomography displacement pulses, we use convex optimization to reconstruct the density matrix $\rho$, shown in Fig.~\ref{fig4}\textbf{e}. We find a state fidelity $\mathcal{F} = \sqrt{\Tr(\rho_{\rm N00N}\cdot|\rho|)} = 0.748\pm 0.008$ to the ideal $N=2$ N00N state $\rho_{\rm N00N}$, in reasonable agreement with our simulation fidelity of $\mathcal{F} = 0.745$ (see Supplementary). The unwanted phonon emission mentioned above, together with the short mechanical resonator lifetimes, limit our ability to generate higher $N$ N00N states.  

In conclusion, we deterministically entangle two macroscopic mechanical resonators on separate substrates, using two independently-controlled superconducting qubits to synthesize and then analyze multi-phonon states, including high fidelity phonon Bell and N00N states. This platform is scalable, supporting simultaneous  entanglement of larger numbers of mechanical resonators, enabling e.g. the direct synthesis of Greenberger-Horne-Zeilinger (GHZ) and W states, as well as synthetic cat states. Using larger SAW cavities with smaller free spectral ranges, each qubit could access multiple acoustic resonances, opening the possibility for multi-mode quantum information processing with small form-factor acoustic devices (see Extended Data Fig.~4). Our architecture promises further insight into the fundamental science of entangled mechanical systems, as well as an approach to distributed quantum computing. This hybrid quantum system would clearly benefit from increased coherence lifetimes for the SAW resonators, which will be essential for implementing more complex quantum operations \cite{Wang2016, Bild2023, Chamberland2022}. This could be achieved by better material growth \cite{Wollack2021}, different device designs, or possibly lowering the frequency of the SAW resonators \cite{lee2023strong}. 

\clearpage

\bibliographystyle{naturemag}
\bibliography{bibliography}

\clearpage
\subsection*{Supplementary}

\textbf{Device fabrication.}
Each acoustic device is fabricated on a LiNbO$_3$ substrate, which is first cleaned using $80^\circ$C Nanostrip to remove organic contaminants. The transducer and acoustic mirrors are  then fabricated by patterning a single layer of 10 nm thick aluminium using a PMMA bilayer liftoff process. The transducers each have 10 finger pairs with a 180 $\mu$m aperture, with a design pitch of 642 nm for node A and 584 nm for node B, while the acoustic mirror pitches are 667 nm and 605 nm, respectively. The distance between the acoustic mirror pairs are $\sim 75~\mu$m for node A and $\sim70~\mu$m for node B. For the qubit die, we first deposit a 100 nm thick aluminium base layer on the sapphire substrate, patterned with optical lithography followed by a plasma etch. Next, a 200 nm thick SiO$_2$ crossover support is patterned using optical lift-off. The qubit and coupler Josephson junctions are then deposited using a standard Dolan bridge technique with bilayer PMMA, with an angled deposition by electron beam evaporation with an intermediate oxidation step. To create galvanic contacts between the junctions and ground plane, we use a bandage layer process with ion milling. The crossover metalizations are completed together with bandage layer. In the final step, we pattern 5 $\mu$m thick standoffs on the sapphire substrate using photo-definable epoxy. The acoustic and qubit chips are then aligned and flip-chip assembled with spacing defined by the standoffs \cite{Satzinger2019}. 

\textbf{Joint state tomography of two mechanical resonators.}
We perform joint Wigner tomography by applying coherent resonant microwave pulses to resonators $R_A$ and $R_B$, using Gaussian pulses with complex amplitudes $\alpha_j,~j = A, B$, where the mean phonon number in each pulse is $|\alpha_j|^2$, with phase distributed over an origin-centered circle in the complex plane. The corresponding displacement operators are given by $\hat{D}_j(-\alpha_j) = \hat{D}_j^\dag(\alpha_j) = \exp(\alpha_j^\star \hat{a}_j - \alpha_j \hat{a}_j^{\dag}),~j = A, B$ where $\hat{a}_j$ is the phonon destruction operator for resonator $R_j$.

Given an initial joint mechanical resonator density matrix $\rho_{m}$, the displacement pulses generate a displaced density matrix
\begin{equation}\label{Eq1}
    \rho_{D} = \hat{D}_A(-\alpha_A) \hat{D}_B (-\alpha_B) \rho_{m} \hat{D}_A(\alpha_A) \hat{D}_B(\alpha_B).
\end{equation}
Following the displacement pulses, each qubit interacts with its respective mechanical resonator, from which we can establish the diagonal elements of $\rho_{D}$ by fitting the time-dependent two-qubit state population traces as in Fig.~\ref{fig3}. The joint mechanical resonator density matrix $\rho_{m}$ can then be found by inverting Eq.~\ref{Eq1} using convex optimization, while constraining $\rho_{m}$ to be Hermitian, positive semi-definite, and trace of $1$. In the joint resonator density matrix reconstruction for the N00N state, we assume a maximum of two excitations in each resonator, so we zero-pad $\rho_{m}$ for phonon indices larger than 2 (note we do not limit the total number of excitations in both resonators).

For the analyses in Figs.~\ref{fig3} and \ref{fig4}, we use displacement pulses distributed over a circle in the complex plane, $\alpha_{j,k} = |\alpha_j| \exp{i 2\pi k / N }, ~j = A, B, ~k=0,1,\ldots, N-1$. For analyzing the two-resonator Bell state, we use $|\alpha_A| = 0.35$ ($|\alpha_B|= 0.26$) with $N=15$, for a total of $15 \times 15 = 225$ pulse combinations. For the N00N state analysis, we use $|\alpha_{A,B}| = 0.3$ with $N=6$, together with $|\alpha_{A,B}| = 0.5$ with $N=15$ for a total of 261 pulse combinations. Uncertainties for the reconstructed density matrices are calculated using a bootstrap method \cite{Efron1993}, randomly selecting with replacement a subset of the pulse combinations and repeating the reconstruction 10 times.

\textbf{Numerical simulations.}
Our system is well-modeled by the Hamiltonian
\begin{align}\label{eq2}
    H &= \sum_{j=A,B} \left [H_{Q_j} + \omega_{R_j} a_j^\dagger a_j + g_{ge,j} (s_{ge,j}^\dagger a_j + h.c.)
    + g_{ef,j} (s_{ef,j}^\dagger a_j + h.c.)\right ] \\ &+ g_q (s_{ge,A}^\dagger s_{ge,B} +h.c.),
\end{align}
In this Hamiltonian, we model each qubit as a frequency-tunable, three-level anharmonic oscillator, with
\begin{equation}\label{eq3}
    H_{Q_j} = \begin{bmatrix} 0 & 0 & 0 \\ 0 &~~\omega_{ge, j}(t) & 0 \\ 0 & 0 & \omega_{gf, j}(t) \end{bmatrix},
\end{equation}
for $j = A, B$. The mechanical resonators have fixed frequencies $\omega_{R_j}$, $j = A, B$. The coupling strength between each qubit and its respective mechanical resonator is $g_{ge,j}$ and $g_{ef,j}$, depending on whether we are coupling the $g \leftrightarrow e$ or the $e\leftrightarrow f$ qubit transitions, with corresponding qubit operators $s_{ge,j} = |g\rangle \langle e|$ and $s_{ef,j} = |e\rangle \langle f|$. The qubit-qubit coupling strength is $g_q= 8.6$ MHz, and is used for preparing the initial qubit Bell states. We use independently-measured system parameters, as given in Extended Data Table.~1, for the Lindblad master equation simulations, which are performed using the open-source Python package QuTiP \cite{Johansson2012}. We note that during the qubit-mechanical resonator Rabi swaps, the qubit $T_1$ is shortened, probably dominated by unwanted IDT emission outside the mirror bandwidth ($\sim 50$ MHz). Using independently-measured mechanical $T_1^m$ and $T_2^m$, in Fig~\ref{fig2}\textbf{c} we fit the qubit $T_1$ during the $e \leftrightarrow g$ swap.  Simulation results are in good agreement with experimental data.

\textbf{Qubit readout correction.}
Two-qubit measurement corrections \cite{Bialczak2010} are applied to all the qubit measurement data. We measure both qubits simultaneously using a multiplexed readout pulse. Prior to each experiment, we measure the two-qubit readout visibility matrix, by preparing the two qubits in the fiducial states $\qty\big{\ket{gg}, \ket{ge}, \ket{eg}, \ket{ee}}$, followed by a two-qubit readout. The visibility matrix $V$ is defined as the transformation between the measured probability vector ($P_{meas}$) and the expected probability vector ($P_{exp}$) for the different fiducial states, $P_{meas} = V P_{exp}$. A typical visibility matrix is:

\begin{equation*}
V=\begin{pmatrix}
F_{gg,gg} & F_{gg,ge} & F_{gg,eg} & F_{gg,ee}\\
F_{ge,gg} & F_{ge,ge} & F_{ge,eg} & F_{ge,ee}\\
F_{eg,gg} & F_{eg,ge} & F_{eg,eg} & F_{eg,ee}\\
F_{ee,gg} & F_{ee,ge} & F_{ee,eg} & F_{ee,ee}
\end{pmatrix} = \begin{pmatrix}
0.954 &  0.042 &  0.027 &  0.002\\
0.022 &  0.939 &  0.000 &  0.028\\
0.024 &  0.003 &  0.955 &  0.037\\
0.001 &  0.017 &  0.018 &  0.934
\end{pmatrix},
\end{equation*}
where $F_{a,b}$ represents the fidelity of preparing the two-qubit state in $\ket{a}$ and measuring the two-qubit state in $\ket{b}$. By inverting the visibility matrix we obtain the measurement-corrected two-qubit probability vector $P_{corr} = V^{-1} P_{meas}$.

\clearpage
\subsection*{Extended data figures and tables}

\begin{table}[ht]
\begingroup
\setlength{\tabcolsep}{12pt}
\centering
\begin{tabular}{@{}lcc@{}}
\toprule
Qubits & $Q_A$  & $Q_B$ \\
\midrule
Qubit idle  frequency (GHz)    & 3.245   & 3.557\\
Anharmonicity (MHz)  &-207 & -196\\
Intrinsic lifetime $T_1$ ($\mu s$)   & 40.8   & 19.3 \\
Ramsey dephasing time $T_{2,R}$ ($\mu s$)  & 2.7   & 3.0 \\
Readout resonator frequency (GHz)  & 4.435   & 4.486 \\
$\ket{g}$ state readout fidelity  & 0.977   & 0.976 \\
$\ket{e}$ state readout fidelity  & 0.993   & 0.991 \\
$\ket{f}$ state readout fidelity (three state readout)  & 0.923   & 0.929 \\
$\ket{f}$ state lifetime ($\mu s$)   & 10.1   & 10.5\\
$\ket{e}$ state lifetime during $|e0\rangle \leftrightarrow |g1\rangle$ swap ($ns$)~~~~~~   & 784   & 350\\
\toprule
SAW resonators & $R_A$  & $R_B$ \\
\midrule
Resonator  frequency (GHz)    & 3.027   & 3.295\\
Resonator  $T_{1}^{m}$ (ns)    & 380   & 270 \\
Resonator  $T_{2}^{m}$ (ns)    & 709  & 527 \\
\midrule
Qubit-resonator max coupling $g_{ge}$ (MHz) & 5.9   & 7.1\\
Experimental $g_{ef}$ (MHz) & 3.8   & 3.8\\
\botrule
\end{tabular}
\endgroup
\captionsetup{labelformat=empty}
\caption{Extended Data Table.~1: \textbf{Summary of device parameters.} Qubit lifetimes, dephasing times, and readout fidelities are measured at the qubit idle frequency. $\ket{e}$, $\ket{g}$ state fidelities are the probabilities of measuring $\ket{e}$, $\ket{g}$ states when preparing the qubit in the $\ket{e}$, $\ket{g}$ states. When swapping the qubit state $\ket{f}$ into resonator ($|f0\rangle \leftrightarrow |e1\rangle$), the qubit $\ket{e}$ state suffers from non-resonant phonon emission, resulting in a relatively short lifetime (see main text).}
\end{table}

\clearpage

\begin{table}
\begingroup
\setlength{\tabcolsep}{12pt}
\centering
\begin{tabular}{@{}lcc@{}}
\toprule
SAW resonators & $R_A$  & $R_B$ \\
\midrule
SAW wavelength $\lambda_0$ ($\mu$m) & 1.301 & 1.194\\
Cavity length ($\mu$m)  & 74.0 & 69.6\\
Sound speed (m/s) & \multicolumn {2}{c}{3979} \\
LiNbO$_3$ coupling coefficient $K^2$ (\%) & \multicolumn {2}{c}{5.4} \\   
Transducer finger pairs &  \multicolumn {2}{c}{10} \\
Number of lines in mirror & \multicolumn {2}{c}{400} \\
IDT duty cycle & \multicolumn {2}{c}{0.5} \\
Mirror duty cycle & \multicolumn {2}{c}{0.5} \\
Aperture ($\mu$m) & \multicolumn {2}{c}{180} \\
Metal thickness (nm) & \multicolumn {2}{c}{10} \\
IDT reflection coefficient & \multicolumn {2}{c}{$-0.042j$} \\
Mirror reflection coefficient & \multicolumn {2}{c}{$-0.0267j$} \\
\botrule
\end{tabular}
\endgroup
\captionsetup{labelformat=empty}
\caption{Extended Data Table.~2: \textbf{Summary of SAW resonator design and P-matrix model.} We provide the design parameters for each SAW resonator ($R_A$ and $R_B$) and P-matrix model parameters used for numerical results in Fig.~\ref{fig2}{\bf a}.}
\end{table}

\begin{figure}[ht]%
\centering
\includegraphics[width=0.9\textwidth]{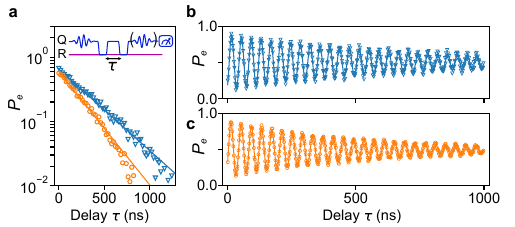}
\captionsetup{labelformat=empty,width=1.0\linewidth}
\caption{
Extended Data Fig.~1: {\bf Mechanical resonator characterization.} Mechanical resonator lifetime ({\bf a}) and coherence time ({\bf b, c}) measurements for resonators $R_A$ (blue) and $R_B$ (orange). Inset in panel \textbf{a} shows the pulse sequence for corresponding experiments (coupler control pulses not shown). Solid lines are least-squares fitting results.}
\label{figS5}
\end{figure}

\begin{figure}[ht]%
\centering
\includegraphics[width=0.8\textwidth]{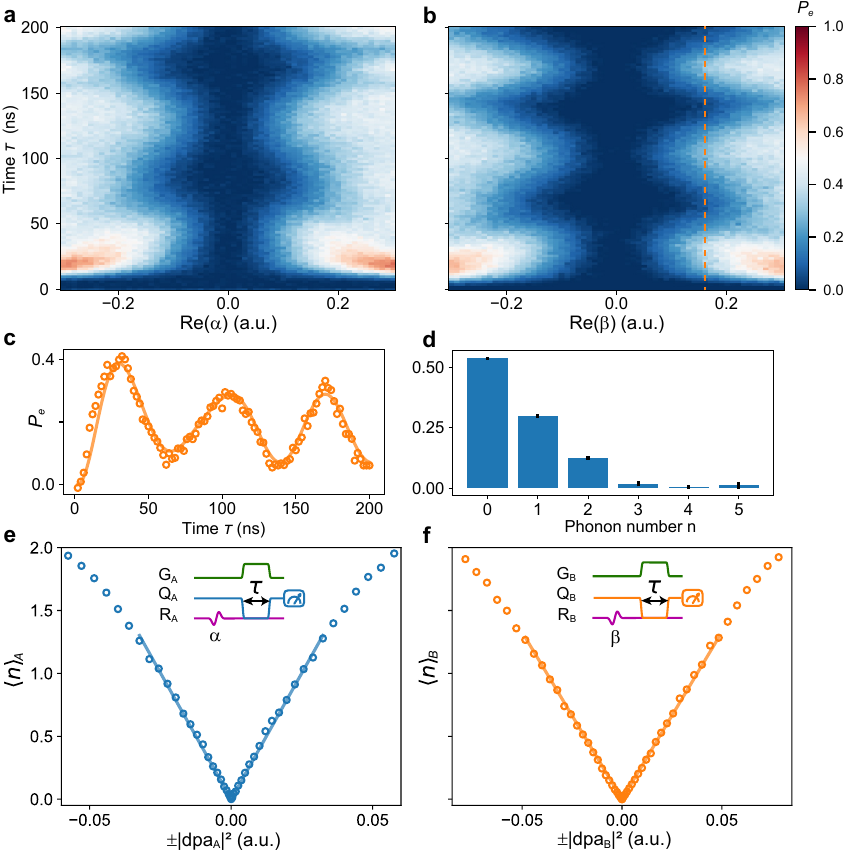}
\captionsetup{labelformat=empty,width=1.0\linewidth}
\caption{
Extended Data Fig.~2: {\bf Displacement pulse calibration.} {\bf a, b,} We displace each mechanical resonator $R_A$ and $R_B$ using a coherent on-resonance drive tone, then measure the average population of the resonator using its associated qubit. The plots show the time-dependent qubit-resonator interactions, for real-only displacement amplitudes $\alpha$ and $\beta$ (horizontal axis) and interaction times $\tau$ (vertical axis). {\bf c, d,} Example fit of the evolution of the phonon number population of resonator $R_B$, from the vertical orange dashed line cut in {\bf b}. The error bars are represented by black vertical lines, indicating one standard deviation. {\bf e, f,} In the low phonon number limit, the average phonon number $\langle n \rangle$ scales linearly with the square of the displacement pulse amplitude (d.p.a). At higher phonon numbers, we attribute the saturation effect to pulse distortion-induced fitting inaccuracy. In the experiment, the displacement pulse amplitudes used are always in the linear range. The average phonon number $\langle n_A \rangle$ and $\langle n_B \rangle$ generated from the respective displacement lines can be estimated from this calibration curve.}\label{figS1}
\end{figure}

\clearpage
\begin{figure}[H]%
\centering
\includegraphics[width=0.9\textwidth]{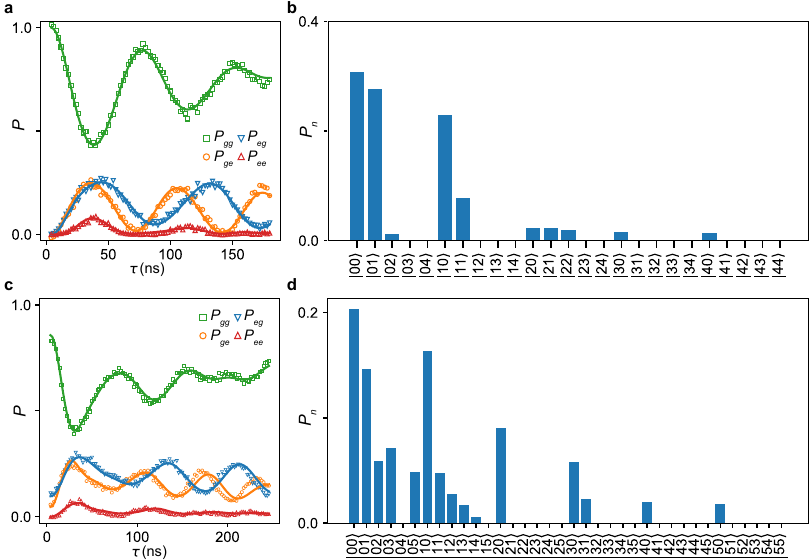}
\captionsetup{labelformat=empty,width=1.0\linewidth}
\caption{Extended Data Fig.~3: \textbf{Wigner tomography for non-zero displacement pulses.} {\bf a, } Qubit-resonator swaps for displacements with $\alpha =0.35$ and $\beta = -0.0271 - 0.258i$ for resonators $R_A$ and $R_B$ respectively. Data are joint qubit measurements. {\bf b,} Fitted probabilities for the corresponding joint mechanical resonator populations. For the fits, we use each resonator's lowest five energy levels. {\bf c, } Qubit-resonator swaps for non-zero displacements with $\alpha =0.5$ and $\beta = 0.5$, here for analysis of the $N=2$ N00N state. {\bf d,} Fit joint mechanical resonator population probabilities, for the data in {\bf c} using six resonator levels in each resonator. For Fig. \ref{fig3} in the main text, we only show three energy levels in each resonator for the Bell state resonator populations, and in Fig. \ref{fig4}, four energy levels in each resonator for the N00N state resonator populations. Here we show all the resonator levels used in the fitting for tomography of the Bell states (5 levels) and N00N states (6 levels). 
}\label{figS2}
\end{figure}
\clearpage

\begin{figure}[H]%
\centering
\includegraphics[width=0.9\textwidth]{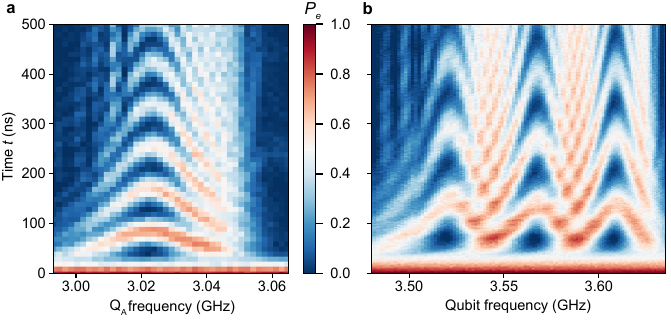}
\captionsetup{labelformat=empty,width=1.0\linewidth}
\caption{
Extended Data Fig.~4: \textbf{Qubit-resonator Rabi swaps with a single mode and a multimode mechanical resonator.} \textbf{a,} Rabi swap measurement with resonator $R_A$, which has a single resonant mode. \textbf{b,} Qubit-resonator Rabi swaps with a different mechanical resonator with three distinct SAW resonances, using a device design similar to that given in the main text. Distance between two acoustic mirrors is about 130$\mu$m, which make the cavity bigger and thus the FSR is decreased to 44 MHz. This illustrates the scalability of this architecture to multi-mode, multi-chip formats.
}\label{figS3}
\end{figure}

\begin{figure}[H]%
\centering
\includegraphics[width=0.65\textwidth]{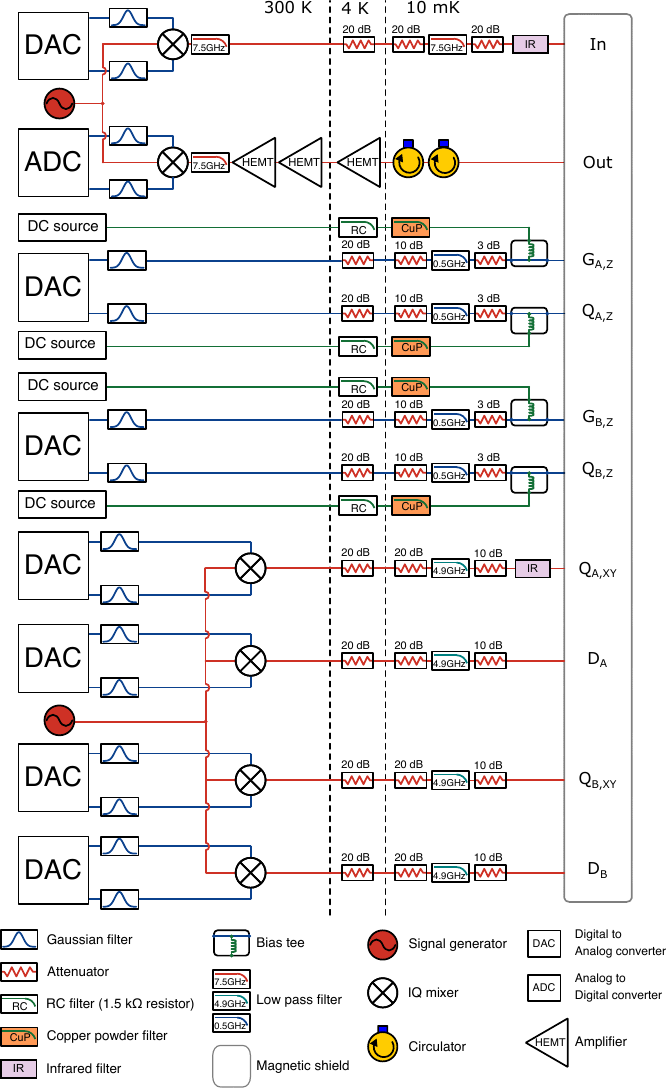}
\captionsetup{labelformat=empty,width=1.0\linewidth}
\caption{
Extended Data Fig.~5: Wiring diagram for the  experiments described in the main text. The input (``In``) and output (``Out``) lines are used for qubit state readout. The qubit control lines include an XY line for exciting each qubit ($Q_{A,XY}$ and $Q_{B, XY}$), and a Z line for changing each qubit's frequency ($Q_{A,Z}$ and $Q_{B,Z}$). The couplers' control lines ($G_{A,Z}$ and $G_{B,Z}$) are used for adjusting the phase of each variable coupler's Josephson junction and thus the coupling strength between each qubit and its associated mechanical resonator. The displacement lines ($D_A$ and $D_B$) are used for applying coherent pulses to each resonator for Wigner tomography.    
}\label{figS4}
\end{figure}

\subsection*{Data availability}
The data displayed in all figures and tables, and other findings of this study, are available from the corresponding author upon reasonable request.

\subsection*{Acknowledgements}
We thank Audrey Bienfait, Youpeng Zhong and Peter Duda for helpful discussions. Devices and experiments were supported by the Air Force Office of Scientific Research (AFOSR grant FA9550-20-1-0270 and AFOSR MURI grant CON-80004392 (GR120272)), DARPA DSO (DARPA agreement HR0011-24-9-0364), and the Army Research Office (ARO grant W911NF2310077). Results are in part based on work supported by the U.S. Department of Energy Office of Science National Quantum Information Science Research Centers. This work was partially supported by UChicago's MRSEC (NSF award DMR-2011854) and by the NSF QLCI for HQAN (NSF award 2016136). We made use of the Pritzker Nanofabrication Facility, which receives support from SHyNE, a node of the National Science Foundation's National Nanotechnology Coordinated Infrastructure (NSF Grant No. NNCI ECCS-2025633). The authors declare no competing financial interests. Correspondence and requests for materials should be addressed to A. N. Cleland (anc@uchicago.edu).

\subsection*{Author contributions}
M.-H.C. designed and fabricated the devices, performed the measurements and analysed the data. H.Q. assisted in measurements and data analysis. H.Y., G.A., and C.R.C. provided suggestions for measurements and data analysis. A.N.C. advised on all efforts. All authors contributed to discussions and production of the manuscript.

\subsection*{Competing interests}
The authors declare no competing interests.

\end{document}